\documentclass[final,1p,times,twocolumn]{elsarticle}
\usepackage{lineno,hyperref,amsmath,amsthm,amssymb,amsfonts,ragged2e,color,subfig}
\journal{``Contributions to Plasma Physics"}
\begin{document}
\begin{frontmatter}
\title{Dust-acoustic envelope solitons in super-thermal plasmas}
\author{A. A. Noman$^{*,1}$, N. A. Chowdhury$^{**,1}$,  A. Mannan$^{1,2}$, and A. A. Mamun$^{1}$}
\address{$^1$ Department of Physics, Jahangirnagar University, Savar, Dhaka-1342, Bangladesh\\
$^2$ Institut f\"{u}r Mathematik, Martin Luther Universit\"{a}t Halle-Wittenberg, Halle, Germany\\
e-mail: $^*$noman179physics@gmail.com, $^{**}$nurealam1743phy@gmail.com}
\begin{abstract}
The modulational instability (MI) of the dust-acoustic waves (DAWs) in an electron-positron-ion-dust
plasma (containing super-thermal electrons, positrons and ions along with negatively charged adiabatic
dust grains) is investigated by the analysis of the nonlinear Schr\"{o}dinger equation (NLSE). To derive the NLSE, the reductive perturbation method has been employed. Two different parametric regions for stable and unstable
DAWs are observed. The presence of super-thermal electrons, positrons and ions significantly
modifies both the stable and unstable regions. The critical wave number $k_c$ (at which modulational
instability sets in) depends on the super-thermal electron, positron, and ion, and adiabatic dust concentrations.
\end{abstract}
\end{frontmatter}
\section{Introduction}
\label{1sec:int}
The existence of positrons in a four components dusty plasma medium (DPM) along with electron-ion-dust
in space plasmas (viz., supernova environments \cite{Jehan2009,Esfandyari-Kalejahi2012,Saberian2016},
in the hot spots on dust rings in the galactic centre \cite{Paul2017}, interstellar clouds \cite{Jehan2009},
pulsar magnetosphere \cite{Esfandyari-Kalejahi2012}, Milky Way \cite{Paul2017}, and accretion disks near
neutron stars \cite{Paul2017}, etc.) as well as laboratory plasmas \cite{Jehan2009} has encouraged researchers
to investigate the propagation of electrostatic perturbation. Plasma physicists have encountered with
dust-acoustic (DA) waves (DAWs) \cite{Jehan2009,Esfandyari-Kalejahi2012,Saberian2016}, dust-ion-acoustic (DIA)
waves (DIAWs) \cite{Paul2017}, DIA cnoidal waves (DIA-CWs) \cite{Saini2016} as well as
their associated nonlinear structures such as shock, solitary \cite{Jehan2009,Esfandyari-Kalejahi2012},
and double layers (DLs) \cite{Alam2014,Dutta2018} to manifest the nonlinear intrinsic properties of
an electron-positron-ion-dust (EPID) plasma medium in presence of highly energetic
positrons \cite{Jehan2009,Esfandyari-Kalejahi2012,Saberian2016,Paul2017}.

The empirical results have shown the signature of the fast particle in the
space \cite{Vasyliunas1968,Saini2016,Alam2014,Dutta2018,Ghosh2012} and laboratory DPM.
These fast particles are governed by the super-thermal/$\kappa$-distribution
which first introduced by the Vasyliunas \cite{Vasyliunas1968}. The $\kappa$-distribution
overlaps with the Maxwellian distribution when the super-thermal parameter $\kappa$ in
$\kappa$-distribution travels to infinite (i.e., $\kappa\rightarrow\infty$). Saini and Sethi \cite{Saini2016}
examined DIA-CWs in a four components DPM, and found that the width and amplitude of the DIA-CWs
decrease with increasing the value of $\kappa$. Alam \textit{et al.} \cite{Alam2014} investigated
solitary waves (SWs) and DLs regarding DIAWs in a DPM in presence of super-thermal electrons, and observed that the amplitude of the
negative DLs potential increases with super-thermal parameter $\kappa$. Dutta and Goswami \cite{Dutta2018} demonstrated
DIA DLs in a four components DPM, and highlighted that the amplitude of the DLs rigourously depends
on the number density of the negative dust as well as super-thermality of the electrons. Ghosh \textit{et al.} \cite{Ghosh2012}
reported  DA SWs (DASWs) in a multi-component plasmas, and found that
the amplitude of the SWs decreases with $\kappa$.

A world of plasma would be a boring place since the nonlinear interaction are at the heart of the working of nature.
The nonlinear Schr\"{o}dinger equation (NLSE) and associated envelope solitonic solutions are the most
interesting theory which can solve the puzzle of nature. A number of authors have investigated the
modulational instability (MI) of the carrier waves by employing the NLSE \cite{Chowdhury2018,Ahmed2018,Bains2013,Saini2008,C1,C2,C3,C4}.
Chowdhury \textit{et al.} \cite{Chowdhury2018} examined the existence of the bright and dark envelope solitons in a
quantum plasma medium, and observed that the thickness of the bright and dark envelope solitons is crucially affected by the
variation of the plasma parameters while the hight of envelope solitons remains constant. Ahmed \textit{et al.} \cite{Ahmed2018}
investigated the MI of the ion-acoustic waves in presence of $\kappa$-distributed electrons and positrons, and found that
the critical wave number ($k_c$) decreases with increasing the value of $\kappa$. Bains \textit{et al.} \cite{Bains2013}
analyzed the  MI of the DAWs in three components DPM. Saini and Kourakis \cite{Saini2008} considered three components DPM having super-thermal plasma species, and investigated the MI of DAWs as well as formation of the envelope solitons, and also found that super-thermality of the plasma species leads to narrower bright envelope solitons.

Recently, a number of authors \cite{Jehan2009,Esfandyari-Kalejahi2012,Saberian2016} have considered a dusty plasma system and studied the nonlinear DASWs. However, to the best knowledge of authors, no attempt has been made on the nonlinear behaviour of DA wave packets and their MI in EPID plasma system. Therefore, in our present work, a four components plasma system (containing super-thermal electrons, positrons, and ions, and adiabatic negatively charged dust grains) has been considered to investigate the nonlinear DAWs and their MI as well as  bright and dark envelope solitons.

The outline of the paper is as follows: The governing equations describing our plasma
model are presented in Section \ref{1sec:gov}. Derivation of NLSE is devoted in
Section \ref{1sec:Derivation of NLSE}. The modulational instability of DAWs is
given in Section \ref{2sec:Modulational instability}. Envelope solitons
are mentioned in Section \ref{2sec:Envelope solitons}. A brief conclusion is,
finally, provided in Section \ref{2sec:Conclusion}.
\section{Governing Equations}
\label{1sec:gov}
We consider a four components plasma model consisting of inertial adiabatic dust grains, inertialess super-thermal
electrons, positrons, and ions. At equilibrium, the quasi-neutrality condition can be
expressed as $ n_{e0}+Z_dn_{d0}=n_{p0}+Z_i n_{i0}$, where $n_{e0}$, $n_{d0}$, $n_{p0}$, and $n_{i0}$ are, respectively,
the equilibrium number densities of electrons, adiabatic dust grains, positrons, and ions.
The normalized governing equations to study the DAWs can be written as:
\begin{eqnarray}
&&\frac{\partial n_d}{\partial t}+\frac{\partial}{\partial x}(n_du_d)=0,
\label{1eq:1}\\
&&\frac{\partial u_d}{\partial t}+u_d\frac{\partial u_d}{\partial x}+\sigma_1n_d\frac{\partial n_d}{\partial x}=
\frac{\partial\phi}{\partial x},
\label{1eq:2}\\
&&\frac{\partial^2\phi}{\partial x^2}=(\sigma_2+\sigma_3-1)n_e-\sigma_2n_p+n_d-\sigma_3n_i,
\label{1eq:3}
\end{eqnarray}
where $n_d$ is the adiabatic dust grains number density normalized by its equilibrium value $n_{d0}$;
$u_d$ is the dust fluid speed normalized by the DA wave speed $C_d=(Z_dk_BT_i/m_d)^{1/2}$
(with $T_i$ being the ion temperature, $m_d$ being the dust grain mass, and $k_B$ being the Boltzmann
constant); $\phi$ is the electrostatic wave potential normalized by $k_BT_i/e$ (with $e$ being the
magnitude of single electron charge); the time and space variables are normalized by
$\omega_{pd}^{-1}=(m_d/4\pi Z_d^2e^2n_{d0})^{1/2}$ and $\lambda_{Dd}=(k_BT_i/4\pi Z_dn_{d0}e^2)^{1/2}$,
respectively; $p_d=p_{d0}(N_d/n_{d0})^\gamma$ [with $p_{d0}$ being the equilibrium adiabatic
pressure of the dust, and $\gamma=(N+2)/N$, where $N$ is the degree of freedom and for one-dimensional case,
$N=1$ then $\gamma=3$]; $p_{d0}=n_{d0}k_BT_d$ (with $T_d$ being the temperatures of the adiabatic dust grains);
and other plasma parameters are considered as $\sigma_1=3T_d/Z_dT_i$, $\sigma_2=n_{p0}/Z_dn_{d0}$, and $\sigma_3=Z_in_{i0}/Z_dn_{d0}$.
The expression for the number density of electrons, positrons, and ions following the
$\kappa$-distribution \cite{Ahmed2018} can be expressed, respectively, as
\begin{eqnarray}
&&n_e=\left[1 -\frac{\sigma_4\phi}{(\kappa_e-3/2)} \right]^{-\kappa+\frac{1}{2}},
\label{1eq:4}\\
&&n_p=\left[1 +\frac{\sigma_5\phi}{(\kappa_p-3/2)} \right]^{-\kappa+\frac{1}{2}},
\label{1eq:5}\\
&&n_i=\left[1 +\frac{\phi}{(\kappa_i-3/2)} \right]^{-\kappa+\frac{1}{2}},
\label{1eq:6}\
\end{eqnarray}
where $\sigma_4=T_i/T_e$ and $\sigma_5=T_i/T_p$. The super-thermality of electrons,
positrons, and ions is, respectively, represented by the $\kappa_e$, $\kappa_p$, and
$\kappa_i$. We consider $\kappa_e=\kappa_p=\kappa_i=\kappa$
for numerical analysis. Now, by substituting \eqref{1eq:4}-\eqref{1eq:6} into \eqref{1eq:3},
and expanding up to third order of $\phi$, we get
\begin{eqnarray}
&&\frac{\partial^2\phi}{\partial x^2}+1=n_d+H_1\phi+H_2\phi^2+H_3\phi^3+\cdot\cdot\cdot,
\label{1eq:7}\
\end{eqnarray}
where
\begin{eqnarray}
&&H_1=(\sigma_2+\sigma_3-1)\gamma_1\sigma_4+\gamma_1\sigma_2\sigma_5+\gamma_1\sigma_3,~~~~~H_2=(\sigma_2+\sigma_3-1)\gamma_2\sigma_4^2-\gamma_2\sigma_2\sigma_5^2-\gamma_2\sigma_3,
\nonumber\\
&&H_3=(\sigma_2+\sigma_3-1)\gamma_3\sigma_4^3+\gamma_3\sigma_2\sigma_5^3+\gamma_3\sigma_3,~~~~~\gamma_1 = (2\kappa-1)/(2\kappa-3),
\nonumber\\
&&\gamma_2 = [(2\kappa-1)(2\kappa+1)]/2(2\kappa-3)^2,~~~~~\gamma_3 = [(2\kappa-1)(2\kappa+1)(2\kappa+3)]/6(2\kappa-3)^3.
\nonumber\
\end{eqnarray}
We note that the term on the right hand side of the Eq. \eqref{1eq:7} is the contribution of
super-thermal electrons, positrons, and ions.
\section{Derivation of NLSE}
\label{1sec:Derivation of NLSE}
To study the MI of DAWs, we will derive the NLSE by employing the standard multiple scale (reductive perturbation) technique \cite{Taniuti1969,Asano2003,Kawahara1973,Jeffrey1982,Selim2015,El-Tantawy2013,Amin1998,Kourakis2003,Kourakis2004,Kourakis2005,Gharaee2011,El-Labany2007,El-Shewy2016,El-Labany2017,Guo2012,Bains2013b,Bouzit2015}. Let A be the state (column) vector $(n_d, u_d, \phi)^T$, describing
the system's state at a given position x and instant t. We shall
consider small deviations from the equilibrium state $A^{(0)}=(1,0,0)^T$
by taking \cite{Kourakis2003,Kourakis2004,Kourakis2005}
\begin{eqnarray}
&&A=A^{(0)}+\epsilon A^{(1)}+\epsilon^2 A^{(2)}+\cdot\cdot\cdot=A^{(0)}+\sum_{n=1}^\infty\epsilon^n A^{(n)},
\label{1eq:8}\
\end{eqnarray}
where $\epsilon\ll 1$ is a smallness parameter. In the standard multiple scale (reductive perturbation) technique, the stretched (slow) space and time variables are commonly used by many authors \cite{Selim2015,El-Tantawy2013,Amin1998,Kourakis2003,Kourakis2004,Kourakis2005,Gharaee2011,El-Labany2007,El-Shewy2016,El-Labany2017,Guo2012,Bains2013b,Bouzit2015} as follows:
\begin{eqnarray}
&&\xi={\epsilon}(x-v_g t),~~~~~\tau={\epsilon}^2 t,
\label{1eq:9}\
\end{eqnarray}
where $v_g$ is the group velocity in the $x$ direction. We assume that all perturbed states depend on the fast scales via
the phase $\theta_1 = kx - \omega t$ only, while the slow scales enter the argument of the $l^{th}$ harmonic amplitude $A_l^{(n)}$, which is allowed to vary along $x$,
\begin{eqnarray}
&&A^{(n)} = \sum_{l = - \infty}^\infty A_l^{(n)}(\xi, \tau)e^{il(kx - \omega t)}.
\label{1eq:10}\
\end{eqnarray}
The reality condition $A_{-l}^{(n)} = A_l^{(n)*}$ is met by all state variables. According to these considerations, the derivative operators are treated as follows
\cite{Jeffrey1982,Selim2015,El-Tantawy2013,Amin1998,Kourakis2003,Kourakis2004,Kourakis2005,Gharaee2011,El-Labany2007,El-Shewy2016,El-Labany2017}
\begin{eqnarray}
&&\frac{\partial}{\partial t}\rightarrow\frac{\partial}{\partial t}-\epsilon v_g\frac{\partial}{\partial\xi}+\epsilon^2\frac{\partial}{\partial\tau},
\label{1eq:13}\\
&&\frac{\partial}{\partial x}\rightarrow\frac{\partial}{\partial x}+\epsilon\frac{\partial}{\partial\xi}\,.
\label{1eq:14}\
\end{eqnarray}
Now, by substituting \eqref{1eq:8}-\eqref{1eq:14} into  \eqref{1eq:1}, \eqref{1eq:2}, and \eqref{1eq:7}, and
collecting the terms containing $\epsilon$, the first order ($m=1$ with $l=1$)  reduced equations can be expressed as
\begin{eqnarray}
&&\hspace*{-1.3cm}iku_{d1}^{(1)}=i\omega n_{d1}^{(1)},
\label{1eq:15}\\
&&\hspace*{-1.3cm}ik\sigma_1n_{d1}^{(1)}=ik\phi_1^{(1)}+i\omega u_{d1}^{(1)},
\label{1eq:16}\\
&&\hspace*{-1.3cm}n_{d1}^{(1)}=-k^2\phi_1^{(1)}-H_1\phi_1^{(1)},
\label{1eq:17}\
\end{eqnarray}
these equations reduce to
\begin{eqnarray}
&&n_{d1}^{(1)}=\frac{k^2}{\sigma_1k^2-\omega^2}\phi_1^{(1)},
\label{1eq:18}\\
&&u_{d1}^{(1)}=\frac{k\omega}{\sigma_1k^2-\omega^2}\phi_1^{(1)},
\label{1eq:19}\
\end{eqnarray}
we thus obtain the dispersion relation for DAWs
\begin{eqnarray}
&&\omega^2=\frac{k^2}{H_1+k^2}+\sigma_1k^2.
\label{1eq:20}\
\end{eqnarray}
The second-order ($m=2$ with $l=1$) equations are given by
\begin{eqnarray}
&&n_{d1}^{(2)}=\frac{k^2}{\sigma_1k^2-\omega^2}\phi_1^{(2)}-\frac{2ik\omega(kv_g-\omega)}{(\sigma_1k^2-\omega^2)^2} \frac{\partial \phi_1^{(1)}}{\partial\xi},
\label{1eq:21}\\
&&u_{d1}^{(2)}=\frac{k \omega}{\sigma_1k^2-\omega^2}\phi_1^{(2)} +\frac{(kv_g-\omega)(k^2\sigma_1+\omega^2)}{i(\sigma_1k^2-\omega^2)} \frac{\partial \phi_1^{(1)}}{\partial\xi},
\label{1eq:22}\
\end{eqnarray}
with the compatibility condition
\begin{eqnarray}
&&v_g=\frac{\omega^2-(\sigma_1k^2-\omega^2)^2}{k\omega}.
\label{1eq:23}
\end{eqnarray}
The coefficients of $\epsilon$ for $m=2$ and $l=2$ provide the second
order harmonic amplitudes which are found to be proportional to $|\phi_1^{(1)}|^2$
\begin{eqnarray}
&&n_{d2}^{(2)}=H_4|\phi_1^{(1)}|^2,
\label{1eq:24}\\
&&u_{d2}^{(2)}=H_5 |\phi_1^{(1)}|^2,
\label{1eq:25}\\
&&\phi_{2}^{(2)}=H_6 |\phi_1^{(1)}|^2,
\label{1eq:26}\
\end{eqnarray}
where
\begin{eqnarray}
&&H_4=\frac{2H_6k^2(\sigma_1k^2-\omega^2)^2-3\omega^2k^4-\sigma_1k^6}{2(\sigma_1k^2-\omega^2)^3},
\nonumber\\
&&H_5=\frac{H_4\omega(\sigma_1k^2-\omega^2)^2-\omega k^4}{k(\sigma_1k^2-\omega^2)^2},
\nonumber\\
&&H_6=\frac{3\omega^2k^4+\sigma_1k^6-2H_2(\sigma_1k^2-\omega^2)^3}{6k^2(\sigma_1k^2-\omega^2)^3}.
\nonumber\
\end{eqnarray}
Now, we consider the expression for  ($m=3$ with $l=0$) and ($m=2$ with $l=0$), which leads the zeroth harmonic modes. Thus, we obtain
\begin{eqnarray}
&&\hspace*{-1.3cm}n_{d0}^{(2)}=H_7|\phi_1^{(1)}|^2,
\label{1eq:27}\\
&&\hspace*{-1.3cm}u_{d0}^{(2)}=H_8|\phi_1^{(1)}|^2,
\label{1eq:28}\\
&&\hspace*{-1.3cm}\phi_0^{(2)}=H_9|\phi_1^{(1)}|^2,
\label{1eq:29}\
\end{eqnarray}
where
\begin{eqnarray}
&&H_7=\frac{H_9(\sigma_1k^2-\omega^2)^2-2v_g\omega k^3-\sigma_1k^4-k^2\omega^2}{(\sigma_1k^2-\omega^2)^2(\sigma_1-v_g^2)},
\nonumber\\
&&H_8=\frac{v_gH_7(\sigma_1k^2-\omega^2)^2-2\omega k^3}{(\sigma_1k^2-\omega^2)^2},
\nonumber\\
&&H_9=\frac{2v_g\omega k^3+\sigma_1k^2+k^2\omega^2-2H_2(\sigma_1k^2-\omega^2)^2(\sigma_1-v_g^2)}{(\sigma_1k^2-\omega^2)^2(1+H_1\sigma_1-H_1v_g^2)}.
\nonumber\
\end{eqnarray}
Finally, the third harmonic modes ($m=3$) and ($l=1$), with the help
of \eqref{1eq:18}-\eqref{1eq:29}, give a set of equations, which
can be reduced to the following NLSE:
\begin{eqnarray}
&&\hspace*{-1.3cm}i\frac{\partial\Phi}{\partial\tau}+P\frac{\partial^2\Phi}{\partial\xi^2}+Q\mid\Phi\mid^2\Phi=0,
\label{1eq:30}\
\end{eqnarray}
where $\Phi=\phi_1^{(1)}$ for simplicity. The dispersion coefficients $P$ is
\begin{eqnarray}
&&\hspace*{-1.3cm}P=\frac{(kv_g-\omega)(\omega^3-3v_gk\omega^2+3\sigma_1\omega k^2-v_g\sigma_1k^3)-(\sigma_1k^2-\omega^2)^3}{2k^2\omega(\sigma_1k^2-\omega^2)}.
\nonumber\
\end{eqnarray}
The nonlinear coefficient $Q$ is
\begin{eqnarray}
&&\hspace*{-0.70cm}Q=\frac{3H_3(\sigma_1k^2-\omega^2)^2+2H_2(H_6+H_9)(\sigma_1k^2-\omega^2)^2-2\omega k^3(H_5+H_8)-(\sigma_1k^4+k^2\omega^2)(H_4+H_7)}{2 \omega k^2}.
\nonumber\
\end{eqnarray}
It is important to mention few more points: The reductive perturbation technique can
also be used to derive the  Korteweg-de Vries  equation
for describing the evolution of a non-modulated waves, i.e. a bare
pulse with no fast oscillations inside the packet. However, the well
known nonlinear mechanism involved in plasma wave dynamics is
amplitude modulation, which may be due to parametric wave
coupling, interaction between high and low frequency modes or
simply to the nonlinear self-interaction of the carrier waves. The
standard method to study this mechanism adopts a multiple scales
perturbation (also known as reductive perturbation \cite{Taniuti1969,Asano2003}) technique,
which generally leads to a NLSE
describing the evolution of a slowly varying wave packet or envelope.
The wave packet may undergo a Benjamine-Feir type MI under
certain conditions. The MI of wave packets in plasmas acts as a
precursor for the formation of bright and dark envelope solitons.
\begin{figure}[t!]
\centering
\includegraphics[width=80mm]{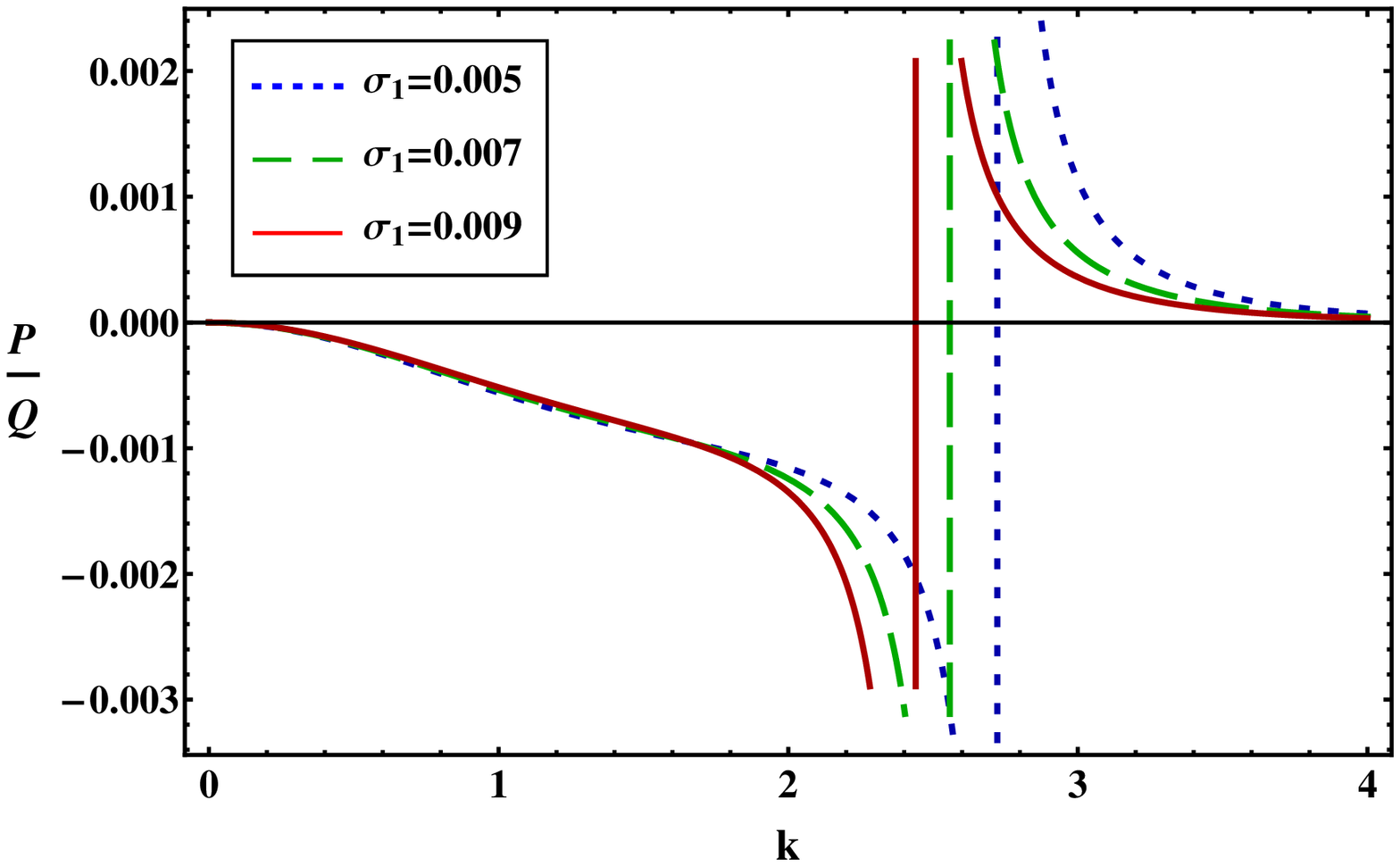}
\caption{Plot of $P/Q$ vs $k$ for different values of $\sigma_1$, when $\sigma_2=0.2$, $\sigma_3=1.3$, $\sigma_4=0.7$, $\sigma_5=0.6$, and $\kappa=1.8$.}
\label{1Fig:F1}
\vspace{0.8cm}
\includegraphics[width=80mm]{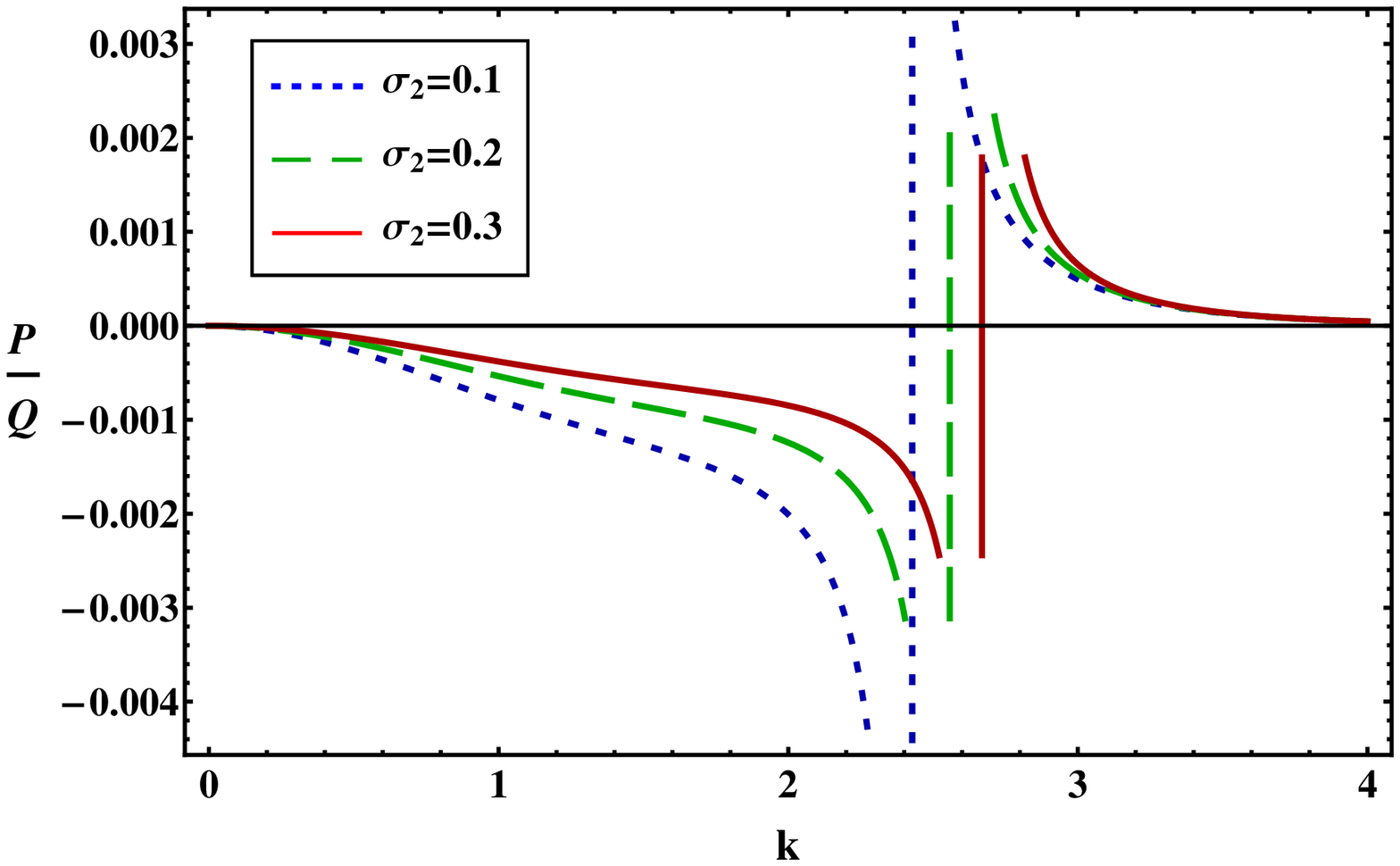}
\caption{Plot of $P/Q$ vs $k$ for different values of $\sigma_2$, when $\sigma_1=0.007$, $\sigma_3=1.3$, $\sigma_4=0.7$, $\sigma_5=0.6$, and $\kappa=1.8$.}
 \label{1Fig:F2}
\end{figure}
\begin{figure}[t!]
\centering
\includegraphics[width=80mm]{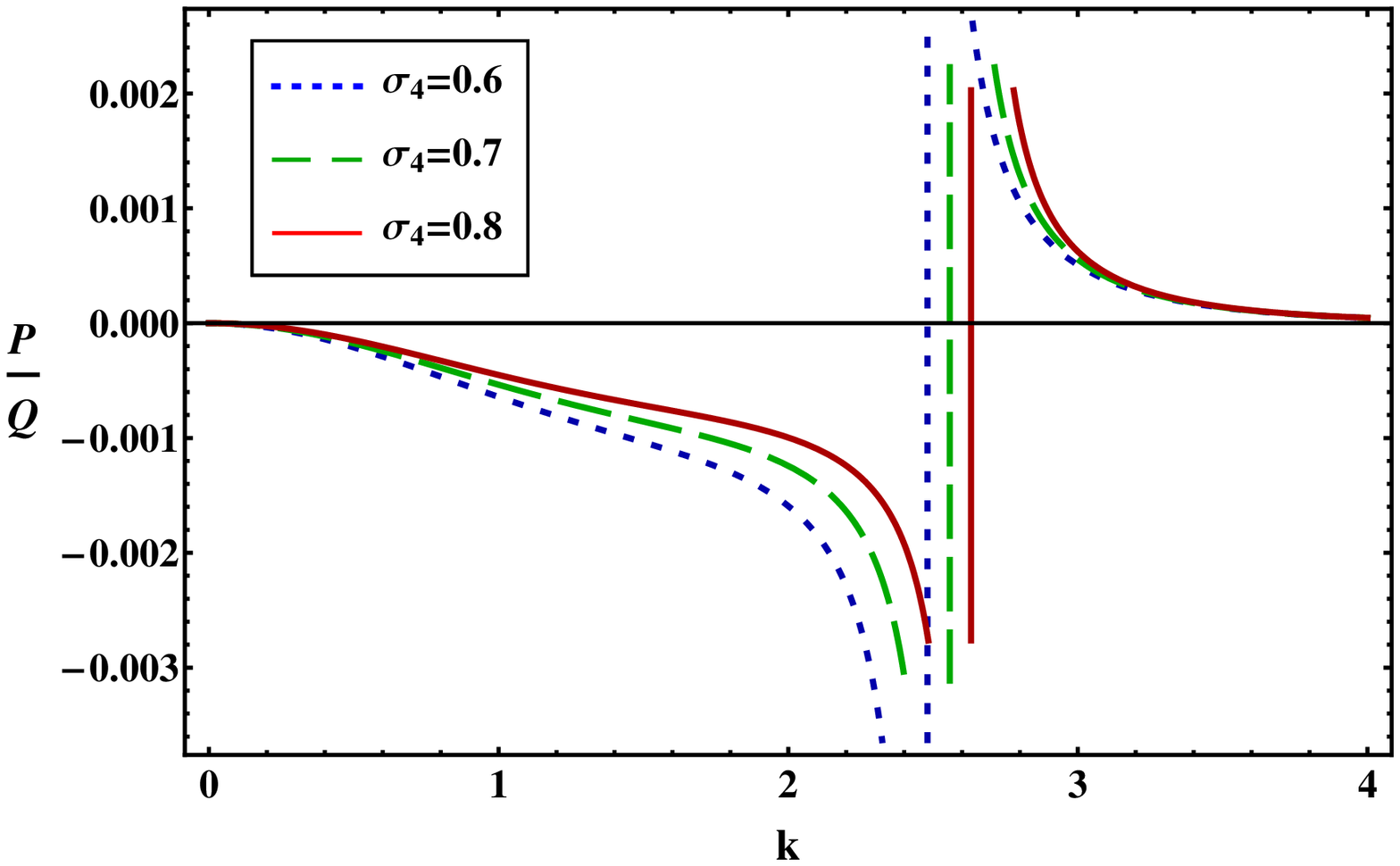}
\caption{Plot of $P/Q$ vs $k$ for different values of $\sigma_4$, when $\sigma_1=0.007$, $\sigma_2=0.2$, $\sigma_3=1.3$,  $\sigma_5=0.6$, and $\kappa=1.8$.}
\label{1Fig:F3}
\vspace{0.8cm}
\includegraphics[width=80mm]{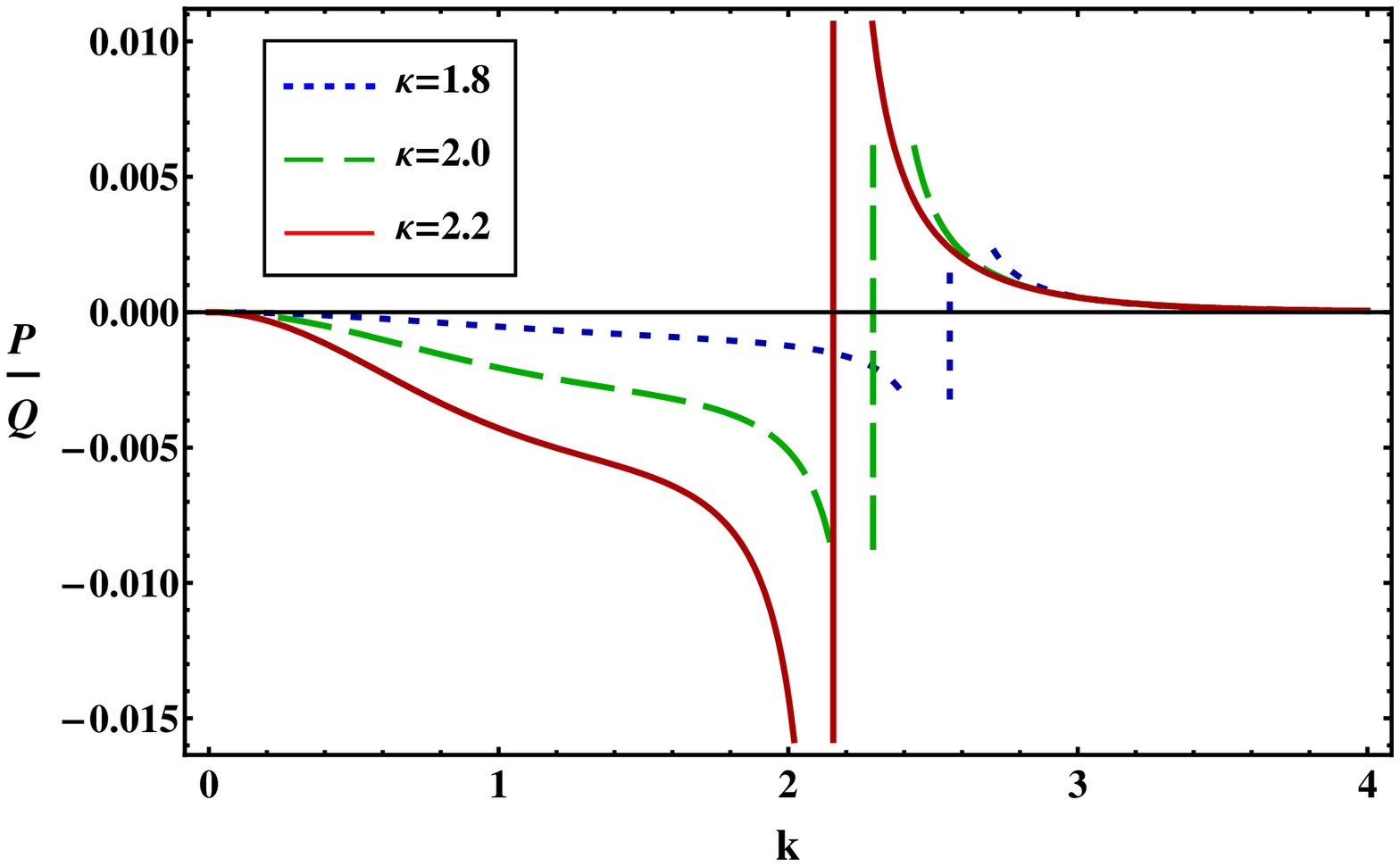}
\caption{Plot of $P/Q$ vs $k$ for different values of $\kappa$, when $\sigma_1=0.007$, $\sigma_2=0.2$, $\sigma_3=1.3$, $\sigma_4=0.7$, and $\sigma_5=0.6$.}
 \label{1Fig:F4}
\end{figure}
\section{Modulational instability}
\label{2sec:Modulational instability}
The stable and unstable domains of the DAWs are organized by the sign of the dispersion ($P$)
and nonlinear ($Q$) coefficients of the standard NLSE \eqref{1eq:30}.
The stability of DAWs in four components DPM is governed by the sign of $P$
and $Q$ \cite{Kourakis2005,Sultana2011,Schamel2002,Fedele2002a,Fedele2002b}. When $P$ and $Q$ are same sign (i.e., $P/Q>0$),
the evolution of the DAWs amplitude is modulationally unstable. On the other hand, when $P$ and $Q$ are
opposite sign (i.e., $P/Q<0$), the DAWs are modulationally stable in presence of the external perturbations.
The plot of $P/Q$ against $k$ yields stable and unstable domains for the DAWs.
The point, at which transition of $P/Q$ curve intersect with $k$-axis, is known as threshold
or critical wave number $k$ ($=k_c$).

The effects of the plasma parameters, specially, the charge state and temperature of the plasma species (via $\sigma_1$)
can be observed in Fig. \ref{1Fig:F1}, and it is obvious that (a) the unstable window, at which MI
sets for the DAWs and allows the formation of the bright envelope solitons, for the DAWs opens for
the large values of $k$ ($k>k_c$); (b) the stable domain for the DAWs can be found for small values of $k$ ($k<k_c$),
and allows the formation of the dark envelope solitons; (c) when $\sigma_1=0.005$, $0.007$, and $0.009$ then
the corresponding $k_c$ value is $k_c\equiv2.45$ (dotted blue curve),
$k_c\equiv2.3$ (dashed green curve), and $k_c\equiv2.2$ (solid red curve);
(d) so, the $\sigma_1$ reduces the stable domain of the DAWs; (e) actually, dust (ion) temperature
reduces (increases) the stable domain of the DAWs for the constant value of dust charge state (via $\sigma_1=3T_d/Z_dT_i$).

Figure \ref{1Fig:F2} indicates the effects of the number density of the positrons and dust grains in recognizing the
stable and unstable domains of the DAWs. It is clear from this figure that (a) the stable domain enhances with the increase in the
value of positron number density whereas negatively charged dust number density suppresses the stable domain for the DAWs when the charge state
of the negative dust remains invariant (via $\sigma_2=n_{p0}/Z_dn_{d0}$); (b) the unstable domain enhances with $Z_d$
for constant value of $n_{p0}$ and $n_{d0}$. So, the number density and the charge state of the negative dust grains
rigourously can change the stable and unstable domains of the DAWs.

The stability domains of the DAWs in four components DPM is also organized by the temperature
of the plasma parameters. Figure \ref{1Fig:F3} describes that the stable (unstable) domain of
the DAWs increases with an increase in the value of the ion (electron) temperature (via $\sigma_4=T_i/T_e$).
The instability criterion of the DAWs for super-thermality of electrons, positrons, and ions in four
components EPID plasma can be observed in Fig. \ref{1Fig:F4}, and it can be seen from this
figure that (a) when for $\kappa=1.8$, $2.0$, and $2.2$ the corresponding $k_c$ value is
$k_c\equiv2.6$ (dotted blue curve), $k_c\equiv2.3$ (dashed green curve), and $k_c\equiv2.2$ (solid red curve);
(b) so, the $k_c$ value decreases with the increase of $\kappa$ and this result is a good agreement with the
result of Ahmed \textit{et al.} \cite{Ahmed2018} work.
\section{Envelope solitons}
\label{2sec:Envelope solitons}
The bright (when $P/Q>0$) and dark (when $P/Q<0$) envelope solitonic solutions, respectively,
can be written as \cite{Kourakis2005,Sultana2011,C6,C8,C9,Schamel2002,Fedele2002a,Fedele2002b}
\begin{eqnarray}
&&\hspace*{-1.4cm}\Phi(\xi,\tau)=\left[\psi_0~\mbox{sech}^2 \left(\frac{\xi-U\tau}{W}\right)\right]^\frac{1}{2}
\times \exp \left[\frac{i}{2P}\left\{U\xi+\left(\Omega_0-\frac{U^2}{2}\right)\tau \right\}\right],
\label{1eq:31}\\
&&\hspace*{-1.4cm}\Phi(\xi,\tau)=\left[\psi_0~\mbox{tanh}^2 \left(\frac{\xi-U\tau}{W}\right)\right]^\frac{1}{2}
\times \exp \left[\frac{i}{2P}\left\{U\xi-\left(\frac{U^2}{2}-2 P Q \psi_0\right)\tau \right\}\right],
\label{1eq:32}
\end{eqnarray}
\begin{figure}[t!]
\centering
\includegraphics[width=80mm]{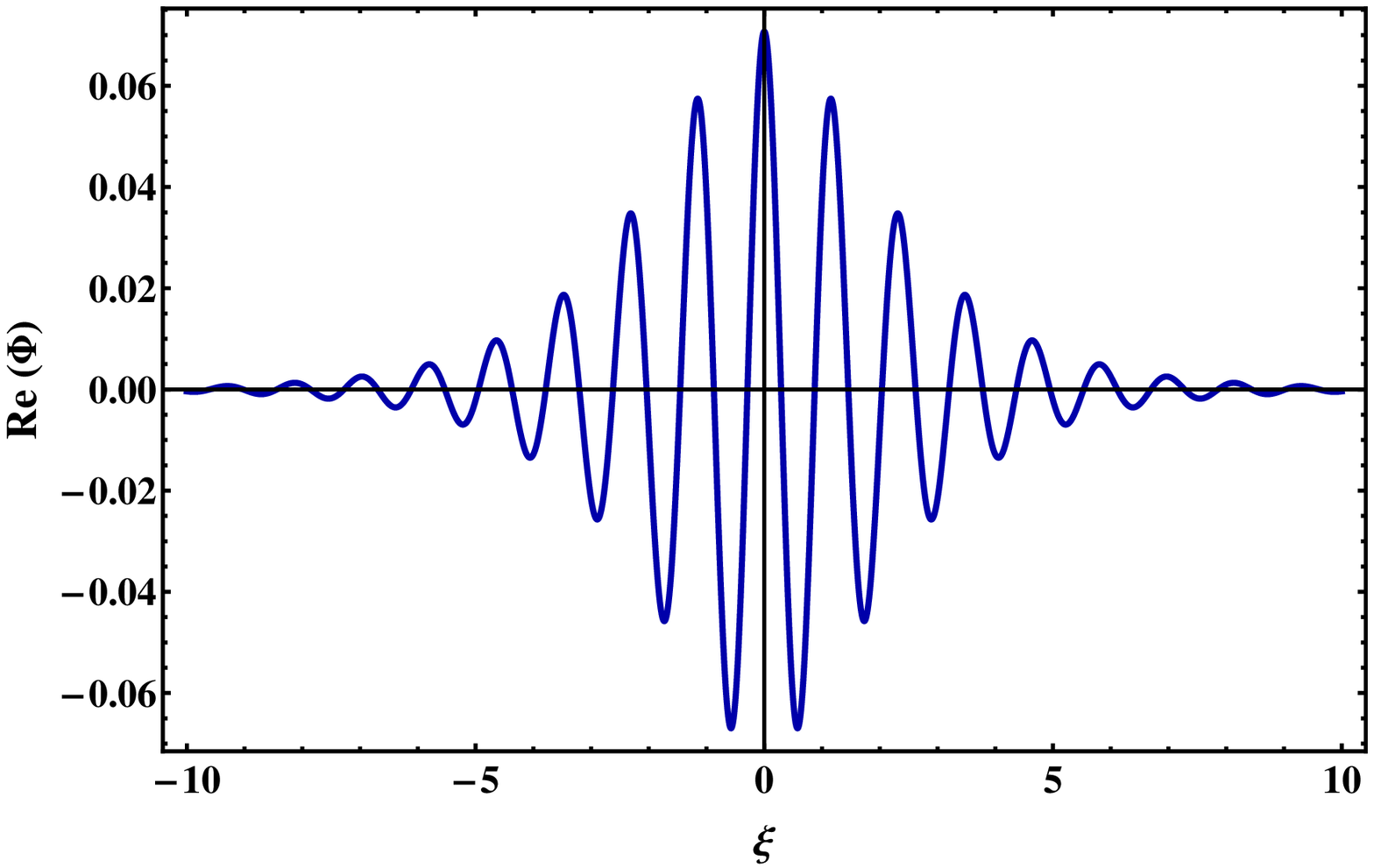}
\caption{Plot of $Re(\Phi)$ vs $\xi$ when $\sigma_1=0.007$, $\sigma_2=0.2$, $\sigma_3=1.3$, $\sigma_4=0.7$, $\sigma_5=0.6$, $\kappa=1.8$, $k=2.5$, $\tau=0$, $U=0.4$, $\Omega_0=0.4$, and $\psi_0=0.005$.}
\label{1Fig:F5}
\vspace{0.8cm}
\includegraphics[width=80mm]{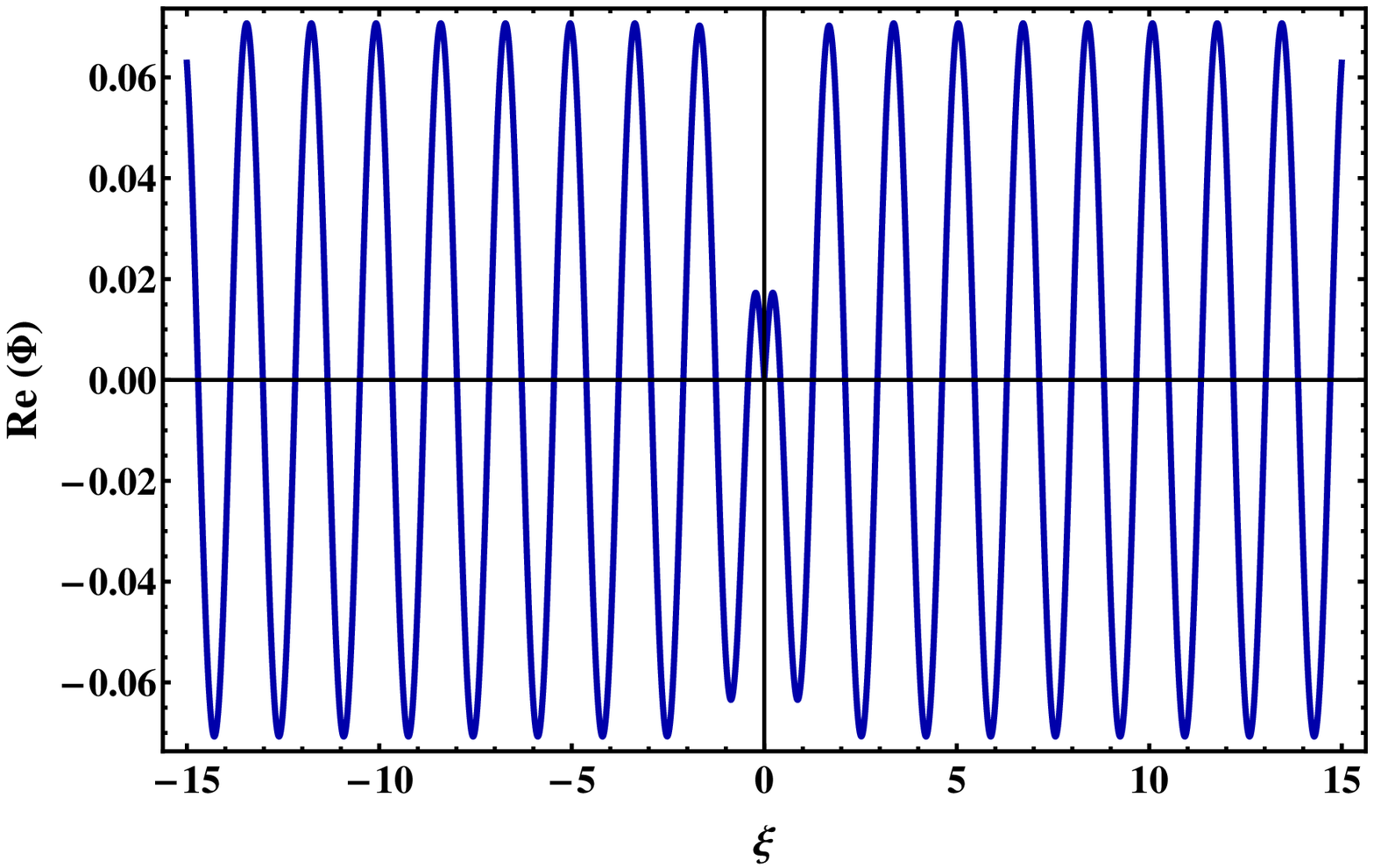}
\caption{Plot of $Re(\Phi)$ vs $\xi$ when $\sigma_1=0.007$, $\sigma_2=0.2$, $\sigma_3=1.3$, $\sigma_4=0.7$, $\sigma_5=0.6$, $\kappa=1.8$, $k=1.5$, $\tau=0$, $U=0.4$, $\Omega_0=0.4$, and $\psi_0=0.005$.}
 \label{1Fig:F6}
\end{figure}
\begin{figure}[t!]
\centering
\includegraphics[width=80mm]{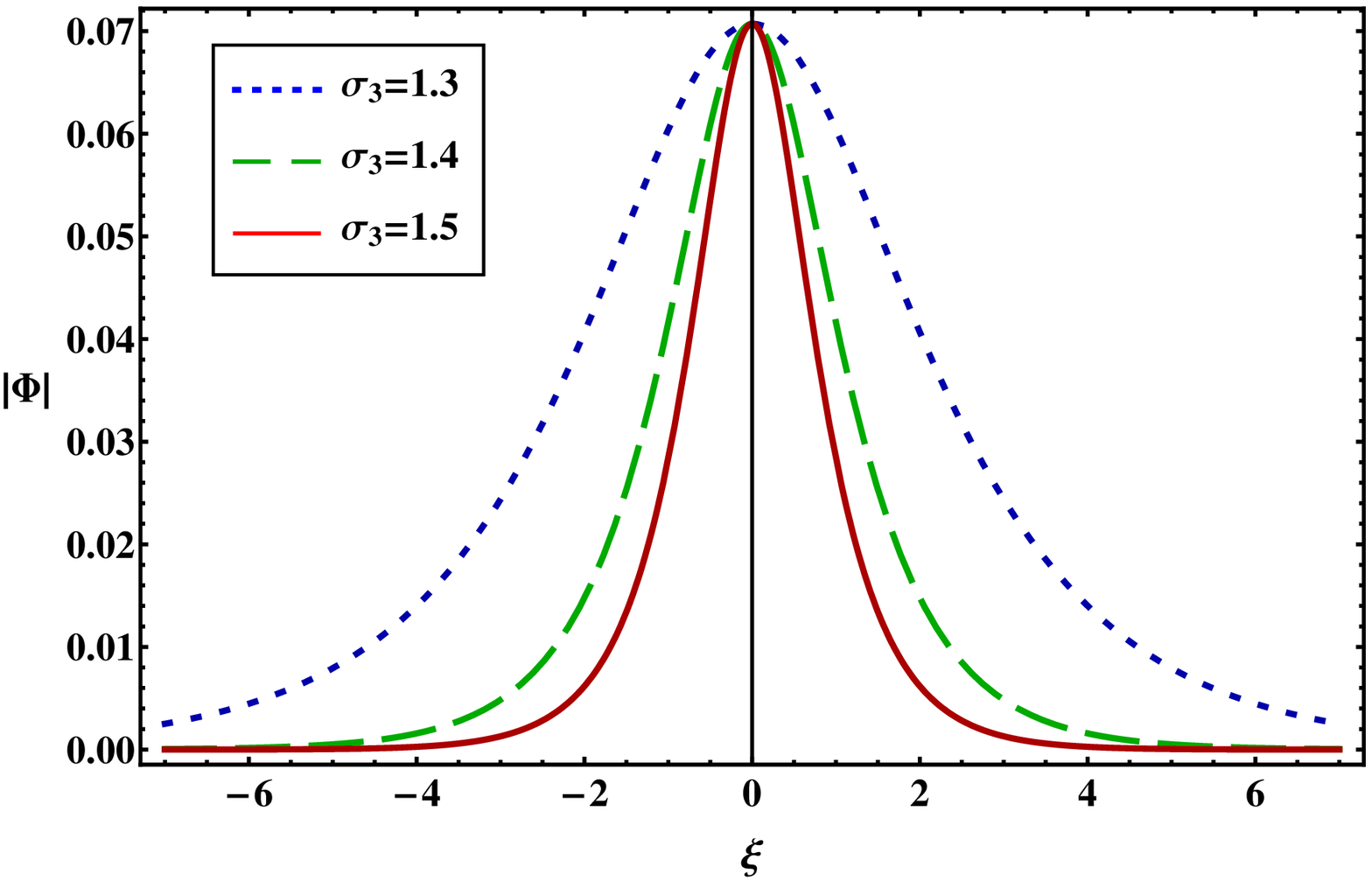}
\caption{Plot of $|\Phi|$ vs $\xi$ for different values of $\sigma_3$, when $\sigma_1=0.007$, $\sigma_2=0.2$, $\sigma_4=0.7$, $\sigma_5=0.6$, $\kappa=1.8$, $k=2.5$, $\tau=0$, $U=0.4$, $\Omega_0=0.4$, and $\psi_0=0.005$.}
\label{1Fig:F7}
\vspace{0.8cm}
\includegraphics[width=80mm]{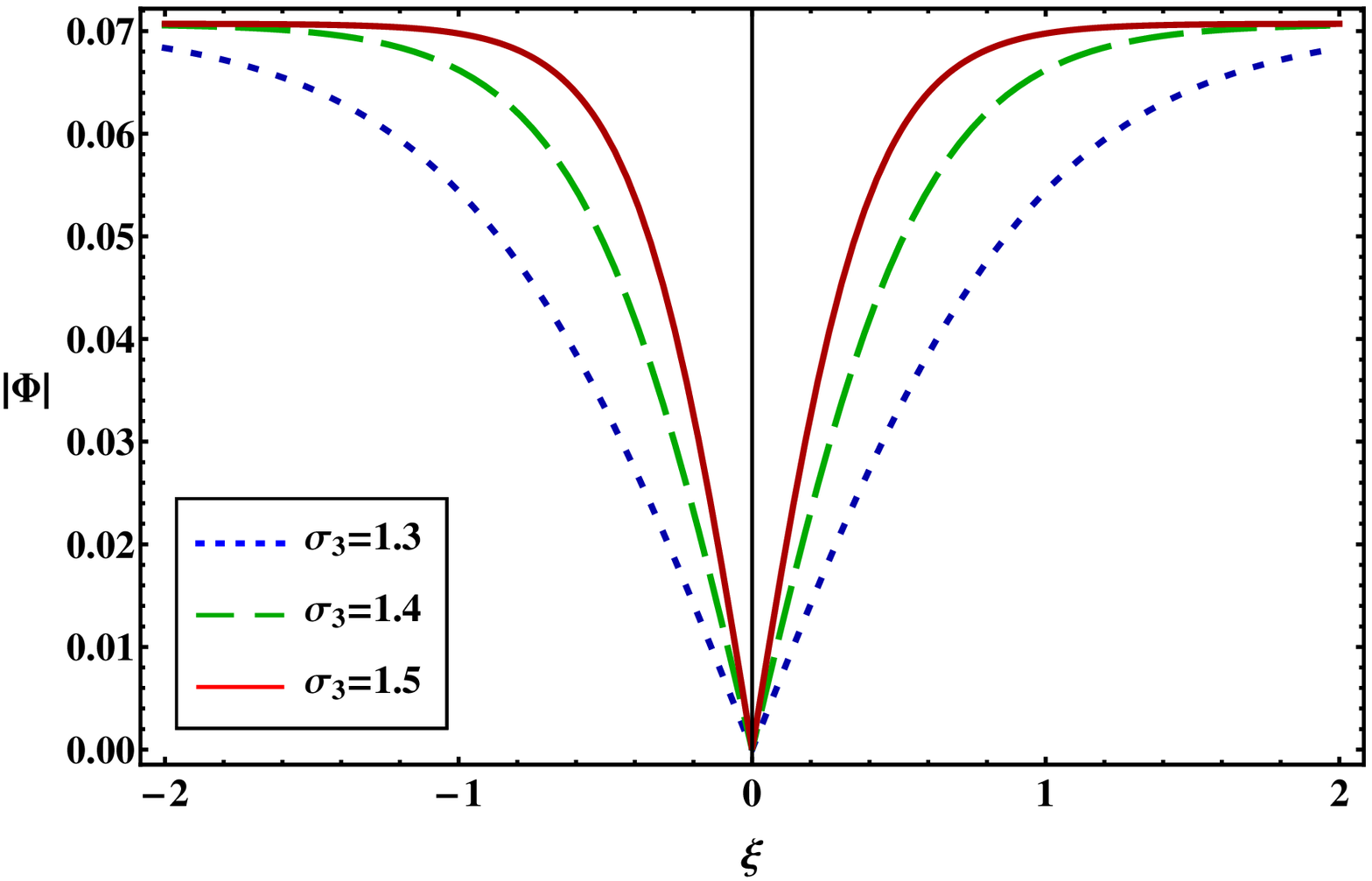}
\caption{Plot of $|\Phi|$ vs $\xi$ for different values of $\sigma_3$, when $\sigma_1=0.007$, $\sigma_2=0.2$, $\sigma_4=0.7$, $\sigma_5=0.6$, $\kappa=1.8$, $k=1.5$, $\tau=0$, $U=0.4$, $\Omega_0=0.4$, and $\psi_0=0.005$.}
 \label{1Fig:F8}
\end{figure}
where $\psi_0$ indicates the envelope amplitude, $U$ is the traveling speed
of the localized pulse, $W$ is the pulse width which can be written
as $W[=(2P\psi_0/Q)^{1/2}]$, and $\Omega_0$ is the oscillating frequency
for $U = 0$. We have depicted bright envelope solitons in Figs. \ref{1Fig:F5} and \ref{1Fig:F7}
by numerically analyzed of equation \eqref{1eq:31}. We have also numerically
analyzed \eqref{1eq:32} in Figs. \ref{1Fig:F6} and \ref{1Fig:F8}. Figure \ref{1Fig:F5}
indicates that (a) the MI of the DAWs associated with unstable domain (i.e., $P/Q>0$)
leads to generate a bare pulse with fast oscillations inside the packet; (b) initially,
the hight of the bare pulse is maximum at $\xi=0$ but decreases with the increase in
the value of positive and negative $\xi$, and still tends to zero for large value of $\xi$ (i.e., $\xi\geq\pm10$).
On the other hand, Fig. \ref{1Fig:F6} describes that (a) the MI of the DAWs associated with
stable domain (i.e., $P/Q<0$) leads to generate a bare pulse with fast
oscillations inside the packet; (b) initially, the hight of the bare pulse is minimum
at $\xi=0$ but increases with the increase in the value of positive and negative $\xi$, and
finally, remains constant for large value of $\xi$ (i.e., $\xi\geq\pm2$).

The nonlinear and dispersion property of the plasma medium as well as the mechanism of the
formation of bright envelope solitons are totally depended on the various plasma
parameters. It is obvious from Fig. \ref{1Fig:F7} that (a) the width of the bright
envelope solitons decreases with the increase in the value of positively charged ion number density (i.e., $n_{i0}$) while
increases with the increase in the value of negatively charged dust grains number density (i.e., $n_{d0}$) when
their charge state (i.e., $Z_i$ and $Z_d$) remain constant (viz., $\sigma_3= Z_i n_{i0}/Z_d n_{d0}$);
(b) the magnitude of the amplitude remain invariant. So, the amplitude is not affected by the variation of the $\sigma_3$.

The influence of the number density of the positively charged ions and negatively charged massive
dust grains as well as their charge state on the formation of dark envelope solitons can be
observed in Fig. \ref{1Fig:F8}, and it is obvious from this figure that  the increase in
the value of $\sigma_3$ causes to change the width of the dark envelope solitons but does not cause any change
in the magnitude of the amplitude of the dark envelope solitons (via $\sigma_3= Z_i n_{i0}/Z_d n_{d0}$).
So, the plasma parameters, viz., $n_{i0}$, $n_{d0}$, $Z_i$, $Z_d$, and $k$
play a vital role in recognising the structure of the dark envelope solitons.
\section{Conclusion}
\label{2sec:Conclusion}
We have investigated the characteristics of the amplitude modulation of DAWs
by using a NLSE, which is successfully derived by employing the standard reductive perturbation technique
in a four components DPM composed of $\kappa$-distributed electrons, positrons,
ions, and negatively charged adiabatic dust grains. In the formation and propagation of DAWs, the moment of inertia is provided by the mass of the adiabatic warm dust grains and restoring force is provided by the thermal pressure of the super-thermal electrons, positrons, and ions.
So, each of the plasma components of the DPM provides a large contribution in the formation and propagation of the DAWs in four components DPM.
A number of authors \cite{Jehan2009,Esfandyari-Kalejahi2012,Saberian2016} have studied DASWs in a four components DPM having electrons, positrons, ions, and negatively charged dust grains. However, they have not studied the nonlinear behaviour of DA wave packets and their MI. Therefore, we have considered this plasma system to investigate the MI of DA wave packet and the formation of the bright and dark envelope solitons in the modulatonally unstable and stable domains of DAWs, respectively. We have also observed that the amplitude of the DA envelope solitons remains constant but the width of the DA envelope solitons changes with the variation of the various plasma parameters. Our results has a good agreement with the previous work \cite{Chowdhury2018}.

To conclude, we hope that our results may be helpful for understanding the MI of DAWs and associated the bright and dark envelope solitons in space plasmas, viz., pulsar magnetosphere, supernova environments, galactic nuclei and also in the laboratory plasmas, viz., intense laser beams.
It may be noted here that the gravitational and magnetic effects are very important to consider but beyond the scope of our present work. In future and for better understanding, someone can investigate the propagation of nonlinear waves in a four components DPM by considering the gravitational and magnetic effects.
\section*{acknowledgements}
The authors are grateful to the anonymous reviewer
for his/her constructive suggestions which have significantly
improved the quality of our manuscript. A. A. Noman is thankful
to the Bangladesh Ministry of Science and Technology
for awarding the National Science and Technology (NST)
Fellowship. A. Mannan thanks the Alexander von Humboldt Foundation for a Postdoctoral Fellowship.

\end{document}